\documentclass[sigconf,screen,review=False]{acmart}

\setcopyright{none}
\renewcommand\footnotetextcopyrightpermission[1]{}
\begin{document}


\title{Feel my Speech: Automatic Speech Emotion Conversion for Tangible, Haptic, or Proxemic Interaction Design}


\settopmatter{authorsperrow=3}

\author{Ilhan Aslan}
\orcid{0000-0002-4803-1290}
\affiliation{%
 \institution{Human-Centered Computing, Aalborg University}
 \country{Denmark}
 }
\email{ilas@cs.aau.dk}

\renewcommand{\shortauthors}{Aslan}

\begin{abstract}

Innovations in interaction design are increasingly driven by progress in machine learning fields. Automatic speech emotion recognition (SER) is such an example field on the rise, creating well performing models, which typically take as input a speech audio sample and provide as output digital labels or values describing the human emotion(s) embedded in the speech audio sample.  Such labels and values are only abstract representations of the felt or expressed emotions, making it challenging to analyse them as experiences and work with them as design material for physical interactions, including tangible, haptic, or proxemic interactions. This paper argues that both the analysis of emotions and their use in interaction designs would benefit from alternative physical representations, which can be directly felt and socially communicated as bodily sensations or spatial behaviours. To this end, a method is described and a starter kit for speech emotion conversion is provided. Furthermore, opportunities of speech emotion conversion for new interaction designs are introduced, such as for interacting with animals or robots. 

\end{abstract}







\maketitle

\section{Introduction}
Emotions in speech can be recognized by analysing both linguistic and paralinguistic features. Arguably, emotion recognition based on paralinguistic features (describing how something is said and not what is being said) is of greater utility and flexibility. One pragmatic reason is that extracting and making use of the linguistic part in speech can be an additional technical challenge and source of error. To transcribe speech into text, an automatic speech recognition (ASR) solution is needed. It is known that the accuracy of ASR can suffer from factors, such as  background noise. The resulting (word) errors impact the performance of the subsequent speech emotion recognition task based on linguistic features \cite{amiriparian2021impact}. Another issue is that the analysis of linguistic features is typically language-dependent and can not be generalized to different languages as easily as paralinguistic features. 

Traditionally, emotion recognition tasks in machine learning (ML) focus on two types of emotion models, predicting (i) emotion labels for discrete emotion categories such as sad, happy, or angry; and predicting values in (ii) continuous emotion dimensions, such as valence and arousal. Arguably, predicting continues dimensions are of greater utility and flexibility. For example, with discrete emotion categories, one is limited to predefined categories. A person who's speech is labelled as happy can be still in very different emotional sates, which fall into the very broad category of happy. Moreover, emotion categories are of limited use if one is interested in studying (continuous) emotion transitions and trajectories \cite{christ2024modeling}.      

Considering the field and task of speech emotion recognition (SER), important progress has been achieved by utilizing transformer architectures, successfully closing the "valence gab" \cite{wagner2023dawn}; i.e., successfully using paralinguistic features only to predict the valence dimension in speech emotion recognition tasks. 

In general, affective computing solutions are becoming increasingly valuable for the design of novel user interfaces that can adapt to users, stimulate positive affective states, and promote well-being. For example, in a previous art installation at Augsburg's architecture museum, in Germany, we explored a proactive conversational design which perceived features of a person based on ML (e.g., age, hair colour, beard, glasses) to generate personalized compliments and deliver them via text and speech synthesis \cite{aslan2023compliment}. In an other work, we had robots telling jokes and adapting their behaviour and joke choices to individual user's humour and affective responses, such as laughter and smiles \cite{weber2018shape}.

Clearly ML based affective computing solutions can be useful in building interactive systems that can perceive complex user behaviours, create novel interactive experiences, and impact users' affective states. But, for various reasons, including technical barriers, it is still challenging to exploit state of the art solutions based on ML for interaction design and artistic research. 

In addition to technical barriers, often there is the issue that research in ML doesn't stop where research in interaction design and human-computer interaction starts. Arguably, ML is about machine-centered research and design, which in many cases has the goal to increase machine capabilities and autonomy. Thus, there is a lack in making such rigid ML models ready to be of flexible use for interaction design and human-computer interaction.  A pragmatic solution today, is to provide tool-support to allow designers in freely and creatively exploring ML models out of their original scope as new design material.

The next section focuses on the topic of converting speech emotions into "physical" emotions (e.g., tangible, haptic, or proxemic experiences). I describe a method and starter kit for automatic speech emotion conversion, and provide an outlook into future design opportunities. For the purpose of this paper, I use the terms emotions or affective states very colloquially.

\begin{figure*}[ht!]
\includegraphics[width= \textwidth]{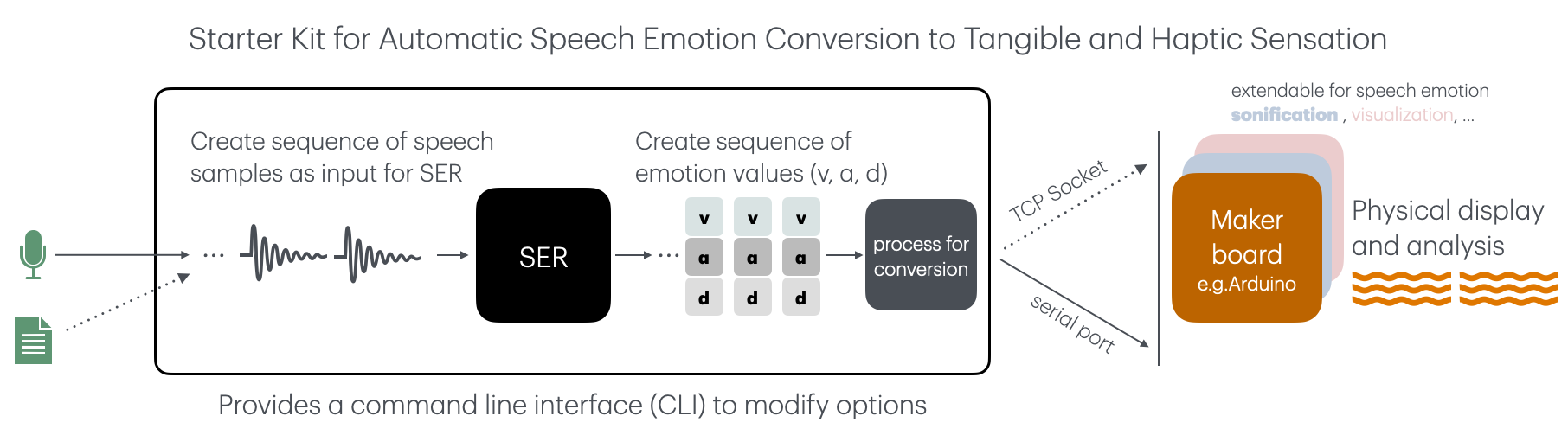}
\caption{Overview of the starter kit. The main part of the kit is provided as a command line tool with a command line interface to easily modify options provided for the processing of the SER task and dealing with specifics of input and output. This part is typically running on a PC. The second part of the toolkit enables exemplary ways to display speech emotions as tangible or haptic sensations with the Arduino IDE.}
\label{fig:overview}
\end{figure*}

\section{"Physical Emotion" Generation from Speech}

One automated way to generate physical emotions from speech is by first using ML models to perform a speech emotion recognition (SER) task and then converting the resulting  emotion values into tangible or haptic sensation and actions with the help of maker kits and the use of servo motors or vibration motors.   

For this approach, a first implementation in the scope of a starter kit is described to help fellow researchers explore physical interaction design ideas utilizing speech emotion conversion (SEC). I plan to make the starter kit available open source.\footnote{\url{https://github.com/islani/SEC_Tool}}

\subsection{Speech Emotion Recognition (SER)}
The field of SER has benefitted from various past emotion recognition challenges, with the first official challenge happening at INTERSPEECH 2009. Since then the field has seen tremendous progress developing new and superior methods especially based on deep neural networks to tackle the task of recognizing emotions in speech. Today, SER is increasingly used to predict continuos emotion dimension values, especially for the dimensions valence, arousal, and dominance. Reflecting on past SER challenges Triantafyllopoulos et al.\ \cite{triantafyllopoulos2024interspeech} describe how SER methods have evolved throughout the last 15 years from using multi-layered perceptrons, long short-term memory recurrent NNs, convolutional NNs, to transformer based models and self-supervised learning approaches. The rapid evolution in SER methods is both impressive and concerning with regards to opportunities for interdisciplinary contributions and application of results in new interaction designs.

The current state of art in SER methods are transformer based \cite{wagner2023dawn} which make use of a wave2vec architecture for self-supervised learning of speech representations \cite{baevski2020wav2vec}, which means that these architectures are typically pre-trained with large amounts of available raw audio data and have already learned to represent audio and speech data in an effective and general way. These pre-trained models can be used as a foundation to create task specific solutions including the task of recognizing emotions in speech taking raw audio as input. Being able to provide SER models speech audio data in raw format without having to preprocess it is really a convenience for anyone who wants to apply SER in interaction designs.  

\subsection{Speech Emotion Conversion (SEC) Kit}
The task of converting speech emotions into physical emotions  can be realized by implementing 3 parts, which need to be interconnected: (i) sensing emotion from speech, (ii) generating tangible, haptic, etc. sensations using physical displays, and (iii) applying rules or methods to map speech emotions into physical sensations and actions. Each part is extendable.

Figure \ref{fig:overview} presents an overview of the system architecture for the proposed starter kit. Intended users of this kit are designers, researchers, educators, or artists who want to exploit speech emotions as design material in their projects and explore novel ideas and bodily experiences. The starter kit integrates the SER part in a command line tool with a command line interface to be flexible and easy to use. Multiple arguments can be set for customization, such as how the continuous speech signal should be divided into audio junks for the SER task, allowing users to receive emotion labels, for example, for every second of audio or every 10 seconds of audio. Different audio junk sizes can result in different emotion labels and depending on use case one may want to explore different sizes. Technically, the speech emotion sensing part is the most demanding part to realize, since it requires state of the art ML solutions and comes with all the challenges of using a black box solution. 

\subsubsection{Realizing affect-sensing}
Fortunately, the number of ready to use affect-sensing tools is increasing. Large IT companies provide  mainly cloud based solutions and APIs for developers to augment their products with affect-sensitive and "affect-exploiting" features. 

In contrast, the AffectToolbox \cite{mertes2024affecttoolbox} is a tool recently provided by researchers of the human-centered AI lab at Augsburg University, which I have had the privilege to be part of in the past. The lab has a long history in providing high quality research on the topic of socially intelligent human-agent interactions and tools to support related research, helping overcome technical tasks, such as signal processing for multimodal human-computer interaction.
The AffectToolBox aims to foster a more inclusive and collaborative environment to advance the field of affective computing research, by being (i) easy to use, (ii) comprehensive, (iii) easy to integrate and (iv) open source. The tool predicts values for continuous emotion dimensions valence(pleasure), arousal, and dominance, allowing the fusion of multiple modalities that are of interest in interactions between humans and embodied agents. For the analysis of human affect, the toolbox provides "out of the box" analysis of paralinguistic features in speech, sentiment analysis (from speech transcriptions), facial expressions and pose from camera. The toolbox is for online analysis and uses live microphone and camera data streams as input. The results are both displayed for the user on a graphical user interface (GUI) and can be send, for example, via UDP to a UDP server to process the results in an other application.   

\begin{figure}[ht!]
\includegraphics[width=\columnwidth]{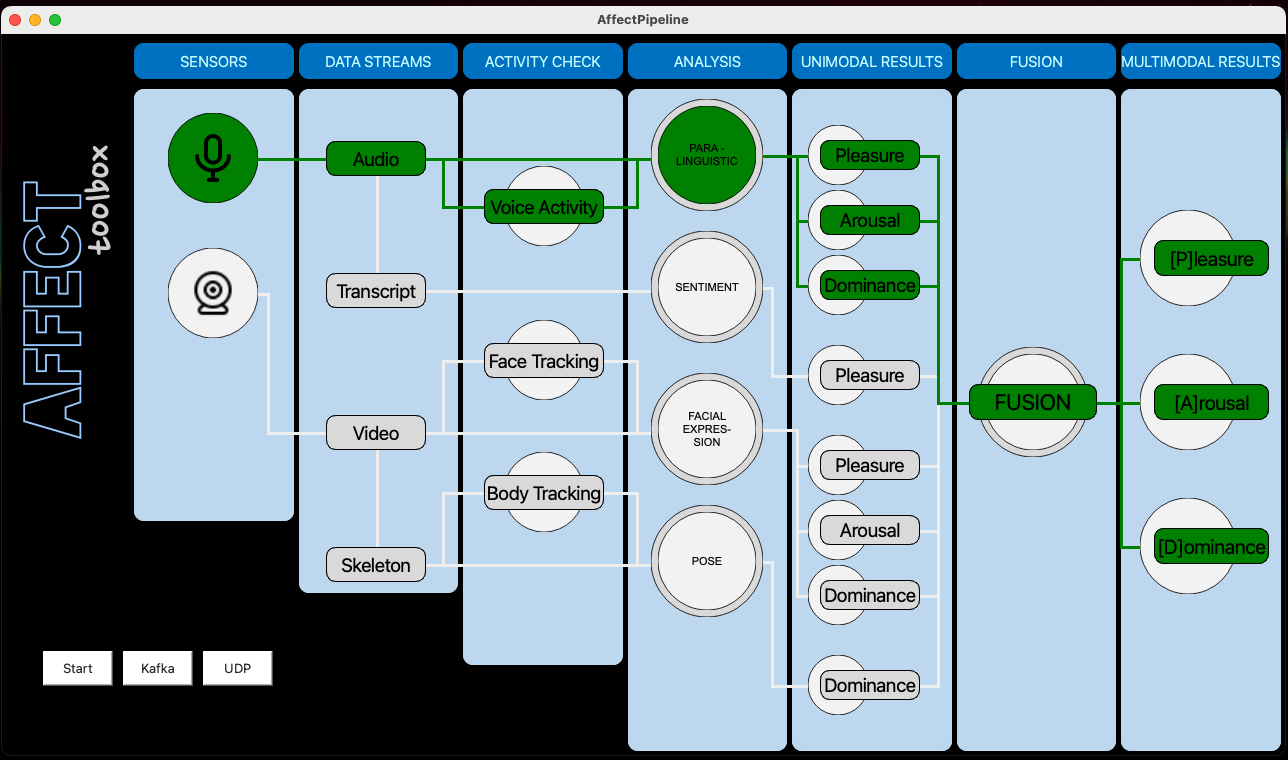}
\caption{Screenshot of the AffectToolbox \cite{mertes2024affecttoolbox} GUI, with  modules selected to perform SER only.}
\label{fig:toolbox}
\end{figure}

The GUI (see Figure \ref{fig:toolbox}) provides users ease of use for selecting the components in the signal processing pipeline they want to use for their specific use case. Considering SER, the Affectoolbox integrates the SER model described and provided by Wagner et al.\ \cite{wagner2023dawn} (another former member of the Human-Centered AI lab in Augsburg) ushering a new era for SER based on transformer architectures and setting a new state of art for SER. 

Overall the AffectToolbox is highly useful for anyone interested in prototyping and studying affect-sensitive systems, including interaction designers, artists, and educators who otherwise would face technical barriers. Arguably, its main strength is in providing a ready to use tool for multimodal human affect analysis. 

As the purpose of this paper is to provide a SEC solution and explore speech emotions as design material, I extracted and modified the SER part from the toolbox and packed it as a command line tool to suit the concrete purpose of creating a (minimal) starter kit for converting abstract speech emotion representations into physical counterparts that can be directly felt, explored, and utilized for new tangibly, haptically, or proxemicly affective interaction designs.

\subsubsection{Computational Generation of "Physical" Emotions}

The command line SER tool accepts as input arguments to set how the emotion recognition results should be send to another device or app for displaying the emotions as physical sensations and integrating them into multisensory interactive experiences. Currently two options are provided for sending the SER results to an external (maker) board (e.g. Arduino\footnote{\url{https://www.arduino.cc/}}) or/and a separate application (e.g. a Processing\footnote{\url{https://processing.org/}} application to visualize speech emotions). 
The starter kit comes with an example for both (i) an Arduino sketch which reads speech emotions over the serial port connection and displays the values physically using a simple vibrotactile motor and (ii) a Processing sketch which receives the SER results through a TCP socket connection and visualizes them. 

I chose these two platforms, since they are known to be suitable for entry level programming and tinkering with large communities of users, including multimedia artists. They are also often used in introductory programming courses in interdisciplinary teaching programs. The design space to map speech emotions to physical sensations and behaviours is vast and there is no one solution fits all since emotion communication is typically context-sensitive and hardware to display physical emotions are not standardized (yet). To design desired speech to physical emotion mappings we, as designers, could for example perform participatory and human centred design to elicit and model user preferences. We have in the past applied participatory design techniques to elicit contextual interaction preferences for gestural interaction \cite{aslan2018pen+}, I believe that similar techniques are suitable for designing physical and contextual mappings of speech emotions. The novelty and subjectivity of speech emotions as design material would also support the application of self-reflective and auto-ethnographic research approaches.

\section{Design Opportunities created by SEC}
 The SEC research is aimed to complement fellow researchers' efforts in addressing for example multisensory experiences \cite{obrist2015emotions,vi2017not} and studying the relationship between affect, emotions, empathy, and (mid-air) haptics \cite{ju2021haptic, ahmed2016reach, ziat2020effects, obrist2015emotions}. 
 Being able to automatically generate physical emotions from speech has many implications for future interaction designs. To illustrate the potential impact of SEC, I provide next a few initial design ideas and opportunities.

\subsection{SEC for Animal-Computer and Animal-Human Interaction }
Human-animal communication and bonding can be beneficial for mental health \cite{beck2003future, walsh2009human} and address issues of loneliness in humans.  Past research has addressed the creation of empathic spaces and identified the potential for artistic responsible research \cite{paananen2023digital, schneiders2023tas}.
Pet owners (like myself) know the role of emotions in communicating with a pet. For example, it is not unusually for vets to instruct owners to keep speaking with their pets in calm and pleasant tones to keep their dog or cat relaxed during health checks or vaccinations. Owners of such companion animals are known to have strong emotional connections with their animals \cite{hall2004psychological}, considering them often as part of the family and providing them with affection, including speech-based affection.  

Quaranta et al.\ \cite{quaranta2020emotion} have demonstrated that cats use both visual and auditory signals to recognize and (appear to) modulate their behaviour to the valence of the perceived emotions. Similar to human-human interactions, technology has also potential to mediate human-animal communication and bonding. The potential may be even higher since animals may only understand a small set of commands. But, animals like humans can need assistive technology. 

In any case, communication with an animal is in general challenging. However, the tone and sentiment in human voices is something that companion animals seem to be able to recognize and use for adapting their behaviour. Haptic feedback is a modality that is already used for pets, there are for example collars which can provide haptic (or auditive) feedback in an attempt to train animals (e.g., to have a cat stay away from danger or from "playing" with the neighbours birds).

A haptic collar could be an additional design option for a (deaf) cat or dog to experience speech emotions. An alternative, which would provide the pets more self-determination could be a pet blanket which can convert speech emotions to physical feedback (haptic, thermal, or smellable). Both design opportunities expand the interaction space and have the potential to fostering (or disrupt if not carefully designed) human-animal bonding. Owner of pets tend to pet their pets while speaking to them, which is of course only possible when owner and pet are co-located. There is some research showing that remote interaction (e.g., video chat) with pets can be reasonable \cite{golbeck2012pet} and in such situations speech emotion conversion could be used to provide automatic petting like physical sensations. There is also emerging research in recognizing vocal emotions of animals \cite{totakura2020prediction}, which could be used to also assist humans in better understanding the emotional vocal expressions of their companion animals. 

\subsection{SEC for Proxemic Interaction}
Proxemic interaction (e.g., \cite{ballendat2010proxemic, greenberg2011proxemic}) was introduce by Saul Greenberg's research group at the University of Calgary, questioning if it is the new ubicomp. Put differently, if enabling machines to interact with humans in spatio-socially intelligent ways is the future goal of ubiquitous computing research. In this future, as one can partly witness today computing is integrated in environments, comes in different physical forms, and interaction designs are increasingly of implicit nature \cite{bittner2019smarthomes}. One could argue, while recently big steps in AI have been made, especially regarding language technologies, machines are still non-mobile and very reactive when it comes to spatial interactions with humans. The most relevant and related progress is happening arguably in robots and autonomous cars. Allowing mobile (and autonomous) machines to share the same physical space is also a (safety) critical decision. 

If we want machines to support humans in (collaborative) physical tasks then there is need to realize proxemic interactions where machines can also use proactive proxemics, i.e., make the first move. To do this they require affective computing abilities, to foresee human action tendencies and communicate action tendencies themselves through both verbal and non-verbal behaviour. 
In the field of human-robot interaction, even simple head and locomotions movements of robots can be used to design emotionally expressive emotions \cite{tsiourti2017designing}. Now, how could speech emotions be converted into proxemic behaviour (e.g., change in body orientations and distance)? One could for example, set automatically the acceptable distance and path that a social robot can take while passing by a person or a group of persons by converting their speech emotions in to physical distances. For example, person(s) sounding angry could map to a minimum proximity of 2 meters to always keep between robot and person(s), while if the speech sound changes to happy then the distance could shrink to 1 meter. Similar controls could be applied to change in body orientation or gaze directions.  
It is worth noting that proxemics is not only about physical space but essentially it is about socio-physical space, including context factors such as (body) heat and smell that are used to organize space. Thus, there is potential for novel designs mapping speech emotions to experiences of smell, taste, or heat.
\cite{obrist2014opportunities, murer2013loll, wilson2011some}

\subsection{SEC for Soemasthetic Interaction Design and Artistic Research}
There are clear potentials to use speech emotion conversion for soma-based design or somaesthetic interaction design \cite{hook2018designing} which embraces an experiential stance towards interaction design. One of the main purposes of converting speech emotions is to extract and digitalize the "emotional essence" into a malleable material that can easily be mapped through design to various bodily experiences.  Somaesthetic interaction design is a rather new field with a limited number of design supporting tools, such as the "soma bits" \cite{windlin2019soma}.  The general idea is to externalize biological signals such as we did in our previous work with heartbeats, mapping a person's heartbeats to a tangible design which allows the person to reflect on their heartbeats or share them with others for socially augmented experiences \cite{aslan2016hold, aslan2020pihearts}. Speech emotions can be considered biodata and fit also into the concept of getting from biodata to somadata \cite{alfaras2020biodata} to enable both first-person and collaborative encounters in design explorations. 
In general ML technologies for perceiving and generating bodily experiences and behaviours has great  potential to serve as a foundation for tools supporting somaesthetic design practices.

The Affective Bar Piano \cite{ritschel2023affective} is a good example of "artistic research" where speech emotion recognition is used by the piano agent to visual adapt its visual appearance (e.g., smile) and the emotional quality of the music it plays to fit the story and tone of the singer. 
In addition to alignment, estrangement can also be strategy in artistic expressions and research, which we can also apply in  embodied design \cite{wilde2017embodied}. Essentially, with our starter kit we provide a way to feel speech emotions as physical sensations and calibrate the "strangeness or familiarity" of experiences by modifying the speech emotion conversion tool's options, and exploring mappings of speech input to outputs of physical displays.

\section{Conclusion}
The future of SER methods is yet undefined. But, one could argue that today we are witnessing a stagnation with models getting ever larger to a degree where less (academic) research groups can contribute to progress in SER due to the lack of available infrastructure. Furthermore,  foundation models are disrupting the landscape of affective computing and increasingly positioning themselves as task independent alternatives for many affective computing tasks \cite{schuller2024affective}. This is both a good and bad development, bad because there is still room for improvement in affective computing tasks but also good, because a slowed down pace of change provides time for interdisciplinary researchers including HCI researcher to explore the current SER solutions and reflect on the new materials that these solutions can produce for interactive system designs.

To facilitate the explorations of speech emotions as malleable design material requires the reduction of technical barriers and flexible use of digital emotions in new design ideas. To this end, this paper described a method and starter kit to enable speech emotion conversion into "physical" emotions that can be expressed through tangilbe, haptic, or proexmic sensations and designs. It was argued that  alternative representations of emotions, which can be directly felt and communicated as bodily sensations create new design opportunities for various domains, including animal computer interaction and artistic research. I have described some initial opportunities and implications for design but the domains addressed were far from being comprehensive. For example, there is a clear opportunity to augment media experiences such as gaming, VR, AR, watching TV or listening to the radio, where speech emotion conversions can make content accessible and experiences potentially more enjoyable.

\bibliographystyle{ACM-Reference-Format}
\bibliography{main}


\end{document}